\documentclass[prl,aps,showpacs,twocolumn]{revtex4}
\usepackage{graphics,bm}
\usepackage{amssymb,ulem}
\usepackage{epsfig}
\usepackage[usenames]{color}

\begin{document}

\title{Critical spin transport in Bose gases}

\author{R. Kittinaradorn}
\affiliation{Institute for Theoretical Physics, Utrecht
University, Leuvenlaan 4, 3584 CE Utrecht, The Netherlands}

\author{R.A. Duine}
\affiliation{Institute for Theoretical Physics, Utrecht
University, Leuvenlaan 4, 3584 CE Utrecht, The Netherlands}

\author{H.T.C. Stoof}
\affiliation{Institute for Theoretical Physics, Utrecht
University, Leuvenlaan 4, 3584 CE Utrecht, The Netherlands}

\date{\today}

\begin{abstract}
We consider spin transport in a two-component atomic Bose gas in three dimensions, at temperatures just above the critical temperature for Bose-Einstein condensation. In these systems the spin conductivity is determined by spin drag, i.e., frictional drag between the two spin components due to interactions.  We find that in the critical region the temperature dependence of the spin conductivity deviates qualitatively from the Boltzmann result and is fully determined by the critical exponents of the phase transition. We discuss the size of the critical region where these results may be observed experimentally.
\end{abstract}


\maketitle

\def\bx{{\bm x}}
\def\bk{{\bm k}}
\def\bK{{\bm K}}
\def\bq{{\bm q}}
\def\br{{\bm r}}
\def\bp{{\bm p}}
\def\half{\frac{1}{2}}
\def\args{(\bm, t)}

{\it Introduction.} --- The research field of spin electronics or spintronics, concerned with practical applications of the electron spin, has renewed interest in spin currents \cite{wolf2001}. In part as a result of these efforts, it is now understood that there are several fundamental differences between charge currents and currents of spin angular momentum. For example, the latter are even under time reversal-symmetry operations and can thus in principle flow without dissipation in ordinary conductors \cite{murakami2003,sinova2004}, contrary to electric currents. Furthermore, the charge conductivity is infinitely large in Galilean invariant systems, whereas spin currents can then still decay due to spin-drag effects, i.e., friction between two spin states due to interactions \cite{scd_giovanni,weber_nature_2005}. Finally, spin and charge current couple in a completely different way to other degrees of freedom in the system, most notably order parameters such as the magnetization or a superconducting condensate. For example, a spin current can exert a so-called spin transfer torque on the magnetization of a ferromagnet \cite{slonczewski1996,berger1996,tsoi1998,myers1999}, a phenomenon that is currently intensively studied in part because of its promise for magnetic-memory applications. On a more fundamental level, there have been several studies on the interplay between spin currents and the critical fluctuations in the magnetization that occur close to the Curie temperature for the ferromagnetic phase transition \cite{fisher1968}. Such magnetic phase transitions form, together with superconducting phase transitions, the overwhelming majority of phase transitions occurring in electronic condensed-matter physics.

In this Letter we consider the effect of another phase transition, i.e., Bose-Einstein condensation, on spin transport. The system we consider is a spin mixture of trapped ultracold bosonic alkali atoms that differs in several important ways from electronic solid-state systems. First, the particles are bosons rather then fermions (electrons). Second, cold-atom systems are disorder free and hence the only contribution to the spin conductivity is the above-mentioned spin-drag effect. The atomic interactions which are responsible for this drag are short ranged as opposed to the Coulomb interactions between the electrons.

\begin{figure}[t]
\begin{center}
\includegraphics[width=1.00\linewidth]{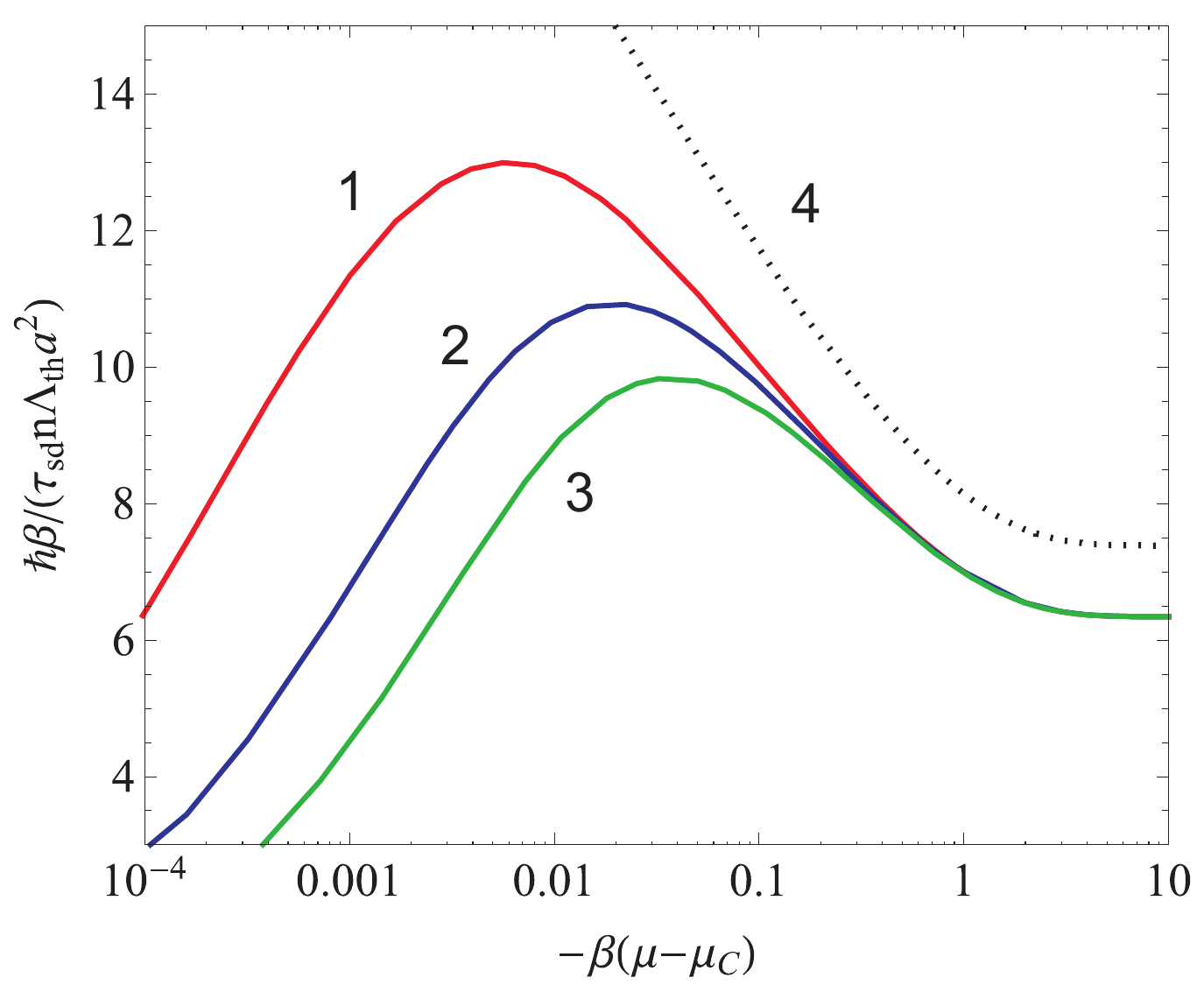}
\caption{(Color online) Spin-drag relaxation rate $1/\tau_{\rm sd}$ at constant temperature as a function of distance from the critical point expressed in terms of the chemical potential difference $\mu-\mu_C$. Lines 1, 2 and 3 represent the relaxation rate for
$a/\Lambda_{\rm th}=9\times10^{-3}$, $2(9\times10^{-3})$ and $3(9\times10^{-3})$, respectively, where $a$ is the scattering length and $\Lambda_{\rm th}=\sqrt{2\pi\hbar^2/(mk_BT)}$ is
thermal de Broglie wavelength. Upon approaching the transition from above, the spin-drag relaxation rate shows an upturn due to Bose enhancement that is ultimately completely suppressed by fluctuations in the critical region. The
dotted line 4 represents the Boltzmann result that does not include critical fluctuations~\cite{driel_prl_2010}. The small quantitative difference between the lines 1-3 and the Boltzmann results far from criticality arises because we neglect vertex corrections in the calculations that lead to the curves 1-3.
\label{fig:main}}
\end{center}
\end{figure}

Spin drag in bosonic cold-atom mixtures was recently studied by two of us using an approach based on the Boltzmann equation. It was found that the bosonic nature of the particles lead to an enhancement of spin-drag effects at low temperatures~\cite{duine_prl_2009}. This should be contrasted with Fermi-liquid behavior that as a result of Pauli blocking leads to suppression of interaction effects at low temperatures so that the so-called spin-drag relaxation rate $1/\tau_{\rm sd} (T)$, which is equal to the inverse of the spin-transport relaxation time $\tau_{\rm sd} (T)$, vanishes quadratically with temperature $T$  \cite{polini_prl_2007,bruun_prl_2008} except in the vicinity of a superconducting \cite{riedl2008} or ferromagnetic phase transition \cite{duine_prl_2010}. For bosons close to the critical temperature for Bose-Einstein condensation, it was found that the Boltzmann approach incorporates the phase transition at the mean-field level and gives $1/\tau_{\rm sd} (T) - 1/\tau_{\rm sd} (T_C) \sim -1/\xi (T) \sim T_C-T$ \cite{driel_prl_2010}, where $\xi (T)$ is the correlation length that diverges at the phase transition.

\begin{figure}[t]
\begin{center}
\includegraphics[width=0.70\linewidth]{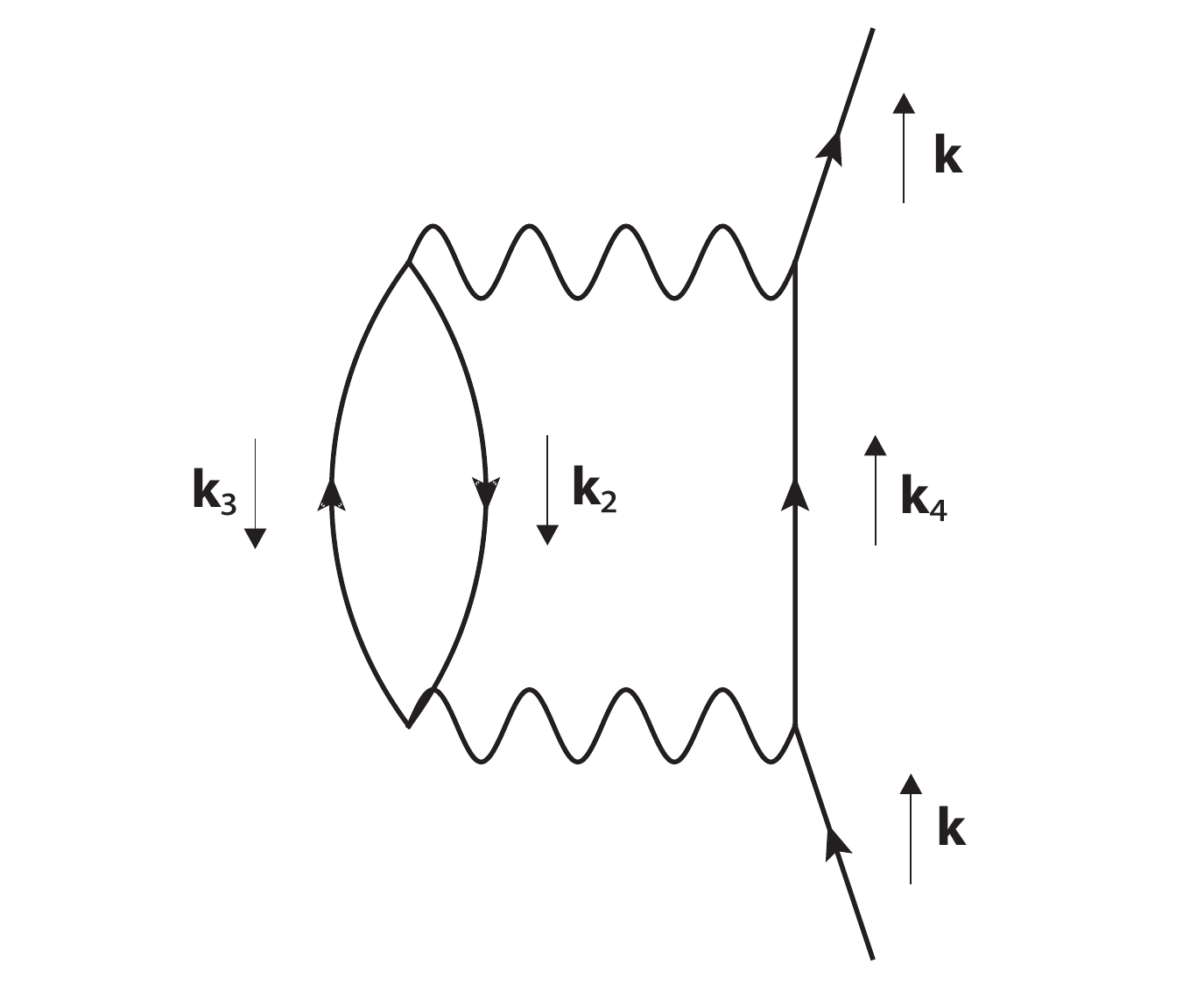}
\caption{Sunset diagram for the self-energy of spin-up particles (the diagram for spin-down is obtained by reversing all spins).
\label{fig:sunset}}
\end{center}
\end{figure}

Our main findings are presented in Fig.~\ref{fig:main} which shows the spin-drag relaxation rate as a function of the distance to the critical point, determined by the difference of the chemical potential $\mu$ from its critical value $\mu_C$. The dotted line shows the Boltzmann result (described above) and the solid lines are the results found using an approach based on the Kubo formula and approximating the atomic self-energy with the so-called sunset Feynman diagram shown in Fig.~\ref{fig:sunset}.  The Hartree diagram should in principle also be included in the self-energy but since this can be achieved by a simple redefinition of the chemical potential we do not consider it here. Within the latter approximation we find that the spin-drag relaxation rate qualitatively agrees with the Boltzmann result for temperatures not too close to the critical temperature, but
in the critical region deviates and goes to zero at the phase transition according to $1/\tau_{\rm sd} (T) \sim 1/\xi (T)$. As we discuss in detail below, an exact scaling {\it ansatz} confirms that the spin-drag relaxation rate vanishes, and in terms of the critical exponents $z$, $\eta$, and $\nu$ we find that $1/\tau_{\rm sd} (T) \sim 1/\xi^{z-d+2-2\eta}$. Below we also discuss the size of the critical region where these effects can be measured.

{\it Spin-drag conductivity.} --- We consider a three-dimensional homogeneous gas of bosonic atoms of mass $m$, with two spin states that couple with opposite sign to an external force $\bm F$. As we discuss in more detail below, this force can in a cold-atom experiment be implemented by a magnetic-field gradient. This force leads to a nonzero spin current ${\bm j}_s$ according to ${\bm j}_s = \sigma_s {\bm F}$, where $\sigma_s=n\tau_{\rm sd}/m$ is the spin conductivity in terms of the spin-drag relaxation time and the density $n$ per spin state (we consider here only the balanced case where the densities of the two spin states are equal). The Kubo formula for the spin conductivity is given by $\sigma_s = -\lim_{\omega\to 0} \mathfrak{Im} \left[ \Pi_{\mu\nu}^{(+)} (\bk={\bm 0},\omega)\right]/\omega$ in terms of the Fourier transform of the retarded spin-current spin-current correlation function $\Pi_{\mu\nu} (\bx,t;\bx',t') = i \theta (t-t')\langle [\hat j^\mu_s (\bx,t),\hat j^\nu_s (\bx',t') ] \rangle/\hbar$, where the expectation value $\langle \cdots \rangle$ is taken in equilibrium and
\[
   j_s^\mu (\bx,t) = \frac{\hbar}{2mi} \sum_{\alpha\in \{ \uparrow,\downarrow\}}\alpha\left[\hat \psi^\dagger_\alpha (\bx,t) \nabla^\mu \hat \psi_\alpha (\bx,t) - {\rm h.c.}\right]~.
\]
In the above expression the bosonic Heisenberg annihilation operator is denoted by $\hat \psi (\bx,t)$ with $\alpha\in\{\uparrow,\downarrow\}$ labeling the spin states, and we note that $\alpha$  takes the respective numerical value $+$ or $-$ if it is not used as a label.

To evaluate the correlation function in the Kubo formula we neglect interactions between atoms of like spins that are of minor importance for spin-drag effects. The hamiltonian then reads
\begin{eqnarray}
&& \hat H = \int d\bx \sum_{\alpha\in \{ \uparrow,\downarrow\}} \hat \psi^\dagger (\bx,t) \left[\frac{-\hbar^2 \nabla^2}{2m}-\mu\right] \hat \psi (\bx,t) \nonumber \\
&&+ T^{2B} \int d\bx \hat \psi^\dagger_\downarrow (\bx,t)\hat \psi^\dagger_\uparrow (\bx,t) \hat \psi_\uparrow (\bx,t)\hat \psi_\downarrow (\bx,t)~,
\end{eqnarray}
with $T^{2B}=4\pi\hbar^2a/m$ the two-body T matrix in terms of the interatomic scattering length $a$ between the two spin states.

Upon ignoring vertex corrections to the correlation function $\Pi_{\mu\nu}^{(+)} (\bk,\omega)$ the Kubo formula for the spin conductivity is worked out to yield
\begin{eqnarray}\label{eq:conductivity}
\sigma_{s} &=& -\frac{\pi\hbar^3}{3m^2}\sum_{\alpha \in\{\uparrow,\downarrow\} } \int\frac{d\bk}{(2\pi)^3}k^2\int d\omega\frac{dN_B(\hbar \omega)}{d\omega}\rho_\alpha^2(\bk,\omega)~,~~~
\end{eqnarray}
with $\rho_\alpha(\bk,\omega)=-\mathfrak{Im} \left[ G^{(+)}_\alpha (\bk,\omega) \right]/\pi\hbar$ the spectral function for particles with spin
$\alpha$, obtained from the Fourier transform of their retarded Green's function that is defined by $G^{(+)}_\alpha(\bx,t;\bx', t') = i \theta (t-t') \left\langle \left[ \hat \psi_\alpha (\bx,t), \hat \psi^\dagger_\alpha (\bx',t')\right]\right\rangle $. Furthermore, $N_B(\hbar\omega)=[e^{\beta \hbar \omega}-1]^{-1}$ is the Bose-Einstein distribution function, with $\beta=1/k_{\rm B}T$ the inverse thermal energy.

The lowest-order (in interatomic interactions) diagram that gives a finite conductivity is the sunset diagram in Fig.~\ref{fig:sunset} for the atomic self-energy
\begin{eqnarray}\label{eq:sunset}
\lefteqn{\mathfrak{Im}[\hbar\Sigma_\alpha^{(+)} (\bk,\omega)] =}\nonumber\\
&& -\pi(T^{2B})^2\int\frac{d\bk_2}{(2\pi)^3}\int\frac{d\bk_3}{(2\pi)^3}\int\frac{d\bk_4}{(2\pi)^3}\nonumber\\
&&\times(2\pi)^3\delta(\bk+\bk_2-\bk_3-\bk_4)\delta(\omega+\mu+\epsilon_{k_2}-\epsilon_{k_3}-\epsilon_{k_4})\nonumber\\
&&\times[N_B(\bk_2)(1+N_B(\bk_3))(1+N_B(\bk_4))\nonumber\\
&&-(1+N_B(\bk_2))N_B(\bk_3)N_B(\bk_4)] ~,~
\end{eqnarray}
with $\epsilon_\bk=\hbar^2k^2/2m$. The imaginary part of the self-energy is calculated numerically from Eq.~(\ref{eq:sunset}). Its real part is obtained by using a Kramers-Kronig relation. The spectral function then follows by using $G_\alpha^{(+)} (\bk,\omega) = \hbar/(\hbar\omega-\epsilon_\bk+\mu-\hbar\Sigma_\alpha^{(+)} (\bk,\omega))$. This real part shifts the critical chemical potential from its noninteracting value of zero to the positive value $\mu_C = \mathfrak{Re} \left[ \hbar \Sigma^{(+)} ({\bm 0}, 0)\right]$.

\begin{figure}[t]
\begin{center}
\includegraphics[width=1.00\linewidth]{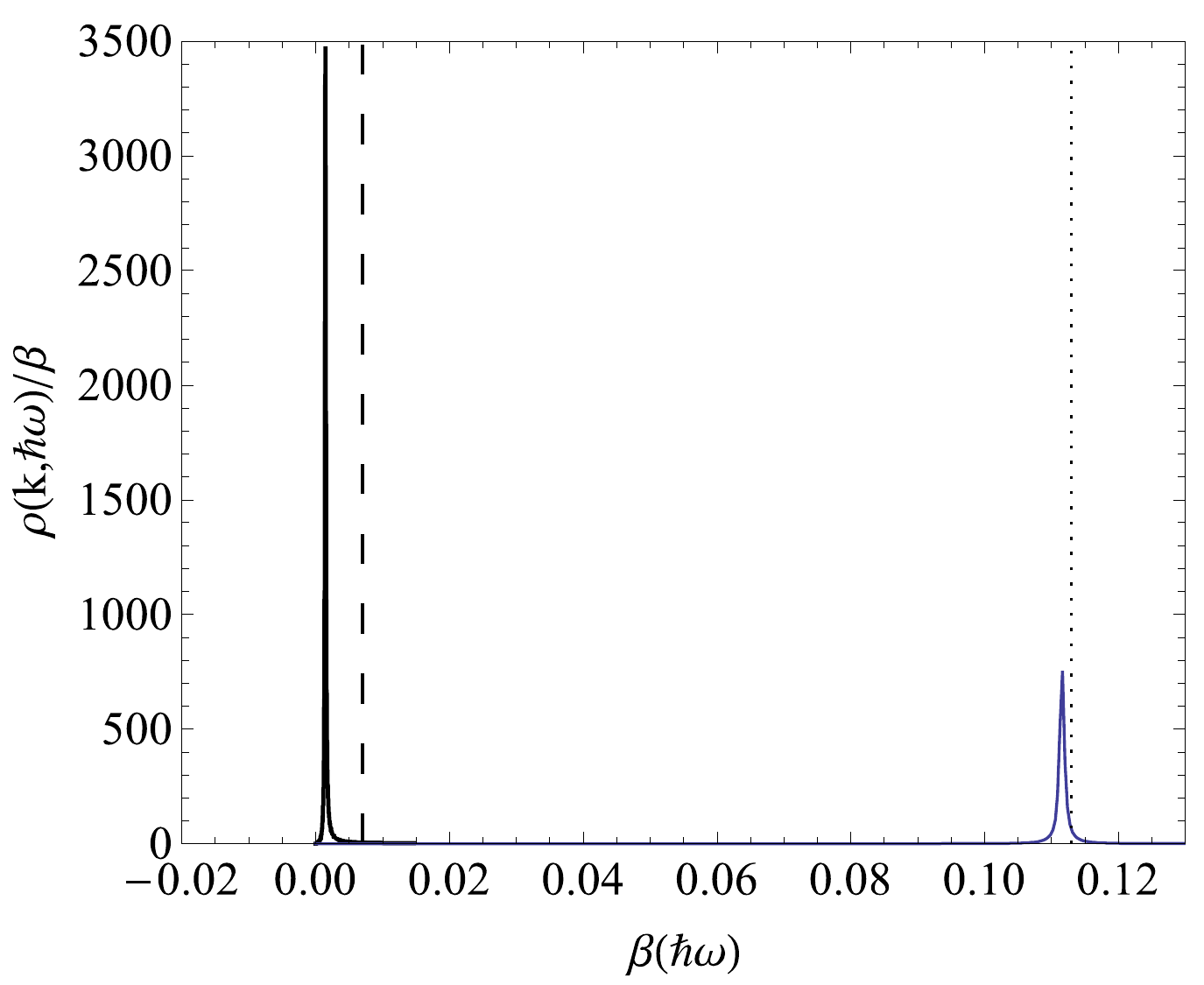}
\caption{The spectral function at low momentum inside (left) and outside (right) the critical region. The dashed and dotted lines show the value of $\epsilon_\bk-\mu$ for the spectral function inside and outside the critical region, respectively.
\label{fig:spectral}}
\end{center}
\end{figure}

The result for the spectral function is shown in Fig.~\ref{fig:spectral} as a function of frequency. For a given momentum, the spectral function exhibits a rather sharp Lorentzian peak corresponding to a quasi-particle excitation. For such a Lorenzian spectral function, the frequency integral in Eq.~(\ref{eq:conductivity}) can be performed and the resulting conductivity is then found to be proportional to the life-time $\tau(\bk)=-\hbar/(2\mathfrak{Im}[\hbar\Sigma^{(+)}(\bk,\omega_\bk)])$ of the quasi-particle, where $\omega_\bk$ is the solution of $\hbar\omega_\bk=\epsilon_\bk-\mu+\mathfrak{Re}[\hbar\Sigma^{(+)}(\bk,\omega_\bk)]$. The evaluation of the expression for the spin conductivity in Eq.~(\ref{eq:conductivity}) with the above expression for the self-energy leads to the results shown in Fig.~\ref{fig:main}
for various scattering lengths.

To gain more insight in the breakdown of the Boltzmann approach in the critical region, we consider the spectral function at low momentum inside (outside) the critical region, corresponding to the left (right) peak in Fig.~\ref{fig:spectral}. The vertical lines correspond to $\epsilon_\bk-\mu$. The peak in the spectral function is shifted considerably from its non-interacting value $\epsilon_\bk-\mu$ in the critical region.
Since the Boltzmann approach does not take into account the shifts in the quasi-particle energy beyond first order in the interaction it does not capture the shift in the critical region correctly. The latter is crucial for the divergence of the conductivity.
The importance of the real part of the atomic self-energy in the critical region is also demonstrated by its importance in determining the upward shift in the critical temperature due to interactions, that is correctly found to be of order $\mathcal{O} \left(a/\Lambda_{\rm th}\right)$
\cite{baym_prl_1999}.

{\it Critical phenomena and scaling.} --- From the calculation that is based on the sunset diagram for the self-energy we find numerically that the spin conductivity diverges as $\sigma_s \sim 1/\sqrt{\mu_C-\mu}$. This result is understood on a more general level by considering scale invariance of the system. Near criticality, we have as a result of the simple (linearized) renormalization-group flow near the fixed point that the spectral function scales as \cite{amitbook}

\begin{equation}\label{eq:scaling}
\rho(\lambda\bk,\lambda^z\omega,\lambda^{1/\nu}(\mu-\mu_C)) = \frac{1}{\lambda^{2-\eta}}\rho(\bk,\omega,\mu-\mu_C)~,~
\end{equation}
where $\lambda$ is an arbitrary dimensionless scaling parameter and $z$, $\nu$, and $\eta$ are critical exponents. The relation between the correlation length $\xi$ and the chemical potential is $\xi\sim 1/|\mu-\mu_C|^{\nu}$. From this scaling {\it ansatz} and Eq.~(\ref{eq:conductivity}) we find that  $\sigma_{s}\sim\xi^{z-d+2-2\eta}\sim|\mu-\mu_C|^{-\nu(z-d+2-2\eta)}$, with $d$ the number of spatial dimensions.

For the sunset diagram in three dimensions, we have $\nu=1/2$, $z=2$, and $\eta=0$, in agreement with the numerical results. It is interesting to note that the behavior of the spin conductivity and the spin-drag relaxation time depends not only on the static critical exponents $\nu$ and $\eta$ but also on the dynamical exponent $z$. We also note that, even though we ignored interactions between atoms with the same spin in our perturbative calculation based on the Feynman diagram in Fig.~\ref{fig:sunset}, the results based on the scaling {\it ansatz} are exact close to the critical temperature and do include these interactions.

Following the reasoning of Hohenberg and Halperin \cite{hhrmp} the factor $\xi^{2-d}$ in these results is understood as follows. A spin-dependent force acting on a region with (fluctuating) spin density $n_s$ is balanced by viscous forces so that $n_s\xi^3{\bm F} \sim \xi^3 \eta_v {\bm v}_s/\xi^2$, where ${\bm v}_s$ is the spin velocity and $\eta_v$ the viscosity. Using that ${\bm j}_s=n_s {\bm v}_s=\sigma_s {\bm F}$ this yields $\sigma_s \sim \xi^2 \langle n_s^2\rangle/\eta_v$. We have that $\langle n_s^2 \rangle \sim \chi_s/\xi^d$ \cite{hhrmp}, with $\chi_s$ the spin susceptibility. In order to obtain full agreement with our result found from the scaling {\it ansatz} we thus need to have that the ratio $\chi_s/\eta_v \sim \xi^{z-2\eta}$. In future work we intend to investigate this conjecture in more detail.

{\it Discussion and conclusions.} --- We have incorporated the effect of critical fluctuations on the behavior of the spin-drag relaxation rate near the critical temperature for Bose-Einstein condensation. We found that the enhancement of the spin-drag relaxation rate due to Bose enhancement of interatomic interactions, predicted by the Boltzmann equation, is suppressed by critical fluctuations sufficiently close to the critical temperature. Numerically, we found the critical region to be proportional to the square of scattering length $|\Delta\mu| \approx60(a/\Lambda_{\rm th})^2$. An estimate based on the Ginzburg criterion \cite{ginzburg} confirms this result. Hence, the size of the critical region may be enlarged by increasing interatomic interactions near a Feshbach resonance. Furthermore, the Ginzburg criterion in $d$ dimensions leads to $|\Delta \mu| \sim (a/\Lambda_{\rm th})^{2/(4-d)}$. Reducing the dimensionality of the system therefore also increases the critical region. With respect to these remarks it is important to note that recent experiments with ultracold bosonic atoms have succeeded in accessing the critical region and measuring the exponent $\nu$ \cite{donner2007}.

The spin conductivity and spin-drag relaxation rate can be measured directly in a drag measurement in which the two clouds of different spin feel a different force due a magnetic-field gradient.  Another method is to study the damping of the spin-dipole mode that is fully determined by the spin-drag relaxation rate.

The main approximation leading to our results is to neglect vertex corrections in the evaluation of the spin-current spin-current response function.  The Boltzmann equation is known to include vertex corrections that essentially lead to a replacement of the single-particle relaxation time by the appropriate transport relaxation time. In the absence of exact cancelations, which we do not expect to occur for the spin-drag conductivity, there is only a quantitative difference between these two time scales, and we attribute the difference between our Kubo approach and the Boltzmann approach sufficiently far away from the critical region to be due to this difference in time scales and hence to be due to neglecting vertex corrections. This assumption is strengthened by noting that far away from the critical region the Kubo and Boltzmann approach have qualitatively the same temperature dependence.

In this work we have considered the effect of thermal critical fluctuations since the phase transition to the Bose-Einstein-condensed state takes place at nonzero temperature. An interesting direction for future work is to investigate also the influence of the vicinity of a quantum critical point \cite{subir} on spin transport in Bose gases, for example by considering the system in an optical lattice where the Mott-insulator-to-superfluid quantum phase transition occurs \cite{greiner2002}. An additional interesting feature of this system is that due to the presence of the optical lattice, which breaks Galilean invariance, now also charge (mass) transport can be considered. 

\acknowledgements This work was supported by a Huygens Scholarship, the Stichting voor Funda-
menteel Onderzoek der Materie (FOM), the Netherlands
Organization for Scientific Research (NWO), and by the
European Research Council (ERC) under the Seventh
Framework Program (FP7). We would like to thank Hedwig van Driel and Vivian Jacobs for their help with the calculations.

\end{document}